\documentclass[fleqn,twoside]{article}
\usepackage{espcrc2}

\usepackage{graphicx}
\usepackage[figuresright]{rotating}

\newcommand{\AmS}{{\protect\the\textfont2
  A\kern-.1667em\lower.5ex\hbox{M}\kern-.125emS}}

\title{The contribution by Domenico Pacini to the Cosmic Ray Physics}

\author{N.Giglietto\address[BA]{Dipatimento Interateneo di Fisica di Bari and
INFN Bari,\\
        via E. Orabona 4, 70126 Bari, Italy}%
        }
       
\begin{document}

\begin{abstract}
Between 1900 and 1913 several people were investigating about the unknown
radiation, later identified as Cosmic Rays. Several experimentalist tried to
identify the origin of this radiation and in particular Victor Franz Hess,
Theodor Wulf and Domenico Pacini. Among them Domenico Pacini had a crucial role
to address the answer to the origin of this radiation in the right way, and 
V.F.~Hess performed the complete set of measurements that definitively excluded
an
origin connected to the soil radioactive elements. 
However the most interesting and may be surprising point it that these
pioneers  defined 1 century ago the three experimental lines to study the
Cosmic Rays: from space, on ground and underground and using only electroscopes.
Domenico Pacini in particular may be considered the pioneer of the underground 
measurements on Cosmic Rays and Hess with his set of systematic measurements
with balloon flights, originated the air-space studies on Cosmic Rays.
\vspace{1pc}
\end{abstract}

\maketitle

\section{Introduction}
This work want to  remember the contribute to the discovery of Cosmic Rays by
Domenico Pacini,
that exactly 1 century ago, around 1910-1911, performed the first underground
measurements of the 
 ''penetrating radiation'', as the Cosmic Rays were called at that time. 

The
measurements by Domenico Pacini were perfectly known in all the world at that
time and cited by several older reviews about the Cosmic
Rays\cite{Wolfendale1984,rossi,hillas,wilson,janossy}
and in particular Wolfendale\cite{Wolfendale1984} noted that the
first one to suggest a non-radioactive origin of Cosmic Rays was Pacini, 
 whose importance is going  to be almost forgotten by latest reviews on this
matter\cite{xu}most of the actual and the
community of people working on Cosmic Rays.

For this reason and approaching  to the european celebrations of the centenary
of the discover of Cosmic Rays, this review, together others that are in
preparation \cite{rob,stra1,stra2,CarlsonDeangelis}, wants to give evidence to
the contribution by Pacini and other
experimentalists that in every case originated the actual ways to study the
Cosmic Rays, and 
that together the set of measurements by V.F.Hess, completed the puzzle about
the origin of this radiation. 
 
\section{First evidences of a unknown radiation}
It was well known since 1785 by Coulomb\cite{Coulomb} that 
electroscopes
spontaneously discharge by the action of the air. After the discovery of the
radioactivity in 1896 by Bequerel\cite{Bequerel} it was well understood that in
presence of radioactive elements charged electroscopes promptly discharged and
the discharge rates were used, at the beginning of the 20th century, to infer
the level of radioactivity. So the logical conclusion about the spontaneuos
discharge of electroscopes in air is that the presence of radioactive elements
on the soil produces charged particles that acts on the electroscopes.

Therefore a huge effort to improve the instruments, understand and identify
the origin and the nature of the unknokn radiation, involved many people during
the beginnings of 1900. In particular,
Wilson, Elster and Geitel \cite{Wilson1901,Geitel} 
modified the electroscope basic
drawings to improve the electroscope insulation and shielding.
 As a result, they could
make quantitative measurements of the rate of spontaneous electroscope
discharge. They
independently concluded that such a discharge was due to ionizing agents coming
from outside
the vessel. 

The obvious questions concerned the nature of such radiation, and
whether it was of terrestrial or extra-terrestrial origin. For example
in 1900 Wilson\cite{Wilson1901} tentatively suggested that the ionizing angent
may be a penetrating radiation of extra-terrestrial origin. 

The simplest
hypothesis however was that its origin was related to radioactive materials,
hence
terrestrial origin was a commonly accepted  assumption.
The experimental demonstration of such hypothesis however was difficult to
probe.

In 1903 Rutherford and Cook \cite{Rutherford} and also McLennan and Burton
\cite{burton} showed that the ionisation was significantly reduced when the
closed vessel was surrounded by shields of metal kept free from radioactive
impurity.  Later
investigations showed that the ionization in a closed vessel was due to a
``penetrating radiation'' partly from the walls of the vessel and partly from
outside.
The lot of measurements by several people at different latitudes are 
summarised by Kurz \cite{kurz} and by
Cline \cite{Cline1910} that discuss about the origin of the radiation.  

Cline\cite{Cline1910} on 1910 summarizes in his paper
the status of art at that moment, and listed the most import measurements
at that time, that were mainly oriented to measure the daily
variations\cite{Borgman1905} or seasonal variation\cite{mach1906}. Cline
in the same paper cited Pacini work \cite{pacini1909} about the daily
variations of the radiation measured on the sea at Sestola in Italy, a
further demonstration of the experimentalist relevance of Pacini at that time. 

This 
Pacini's measurement was particularly remarked in Cline's paper as a first
evidence of the atmosphere as the main responsable of the penetrating
radiation and at the same time excluding the Sun as the main origin.
Cline's conclusions of his own measurements in Canada however,
were that a larger contribution of the radiation should be attributed to the
soil and a negligible contribution from the Sun or
the atmosphere. 

Finally in the review by Kurz\cite{kurz} three possible sources for the
penetrating radiation are discussed: an extra-terrestrial radiation possibly
from the Sun,
radioactivity from the crust of the Earth, and radioactivity in
the atmosphere. Kurz concludes, however, that the possibility of an
extraterrestrial radiation seems to be unlikely. 

It was generally assumed that
large part of the radiation can be accounted as $\gamma$-rays emitted by
radioactive material in the crust\cite{Wul1910,Goc1909}. Calculations
were made of how the radiation should decrease with height (see e.g. Eve
\cite{eve}) and measurements were performed.

\begin{figure}[htb]
\begin{center}
\resizebox{0.75\columnwidth}{!}{\includegraphics{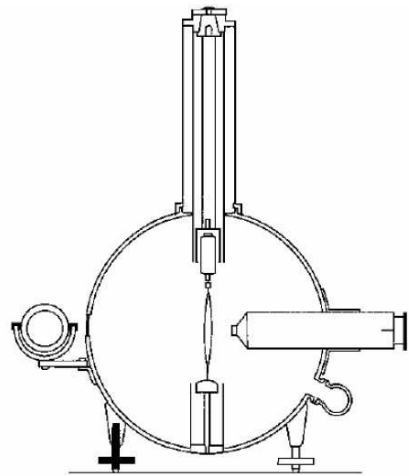} }
\end{center}
\caption{The Wulf electroscope.}
\label{fig:wulf-electroscope} 
\end{figure}

Father Theodor Wulf, a German scientist and a Gesuit priest serving in the
Netherlands and then in Roma,  had the idea to check the variation of
radioactivity with height to test its origin. In 1909 \cite{Wul1910} Wulf, using
an improved  electroscope\cite{Wul1907} (Fig. \ref{fig:wulf-electroscope}) in
which the two leaves had been
replaced by two metalized silicon glass wires, with a tension spring made also
by glass in between, first measured the ionization rate inside a cave at
Valkenburg in Holland then  the rate of ionization at the top of the
Eiffel Tower in 1910 in Paris (300 m above ground).
 Supporting the hypothesis of the
terrestrial origin of most of the radiation, he expected to find at the top a
much smaller ionization than on the ground. The rate of ionization
found, however,  a decrease
 too small  to confirm the hypothesis. He concluded that, in
comparison with the values on the ground, the intensity of radiation ``decreases
at nearly 300 m [altitude] not even to half of its ground value''; while with
the assumption that radiation emerges from the ground there would remain at the
top of the tower ``just a few percent of the ground radiation'' \cite{Wul1910}.

Wulf's observations were of great value, because he could take data at different
hours of the day and for many days at the same place. For a long time, Wulf's
data were considered as the most reliable source of information on the altitude
effect in the penetrating radiation. However Wulf concluded that the most likely
explanation of his puzzling result was still emission from the soil. 

\section{Pacini contribution to the ''unknown radiation'' discussion} 
\begin{figure}[htb]
\includegraphics[height=9cm,width=8cm]{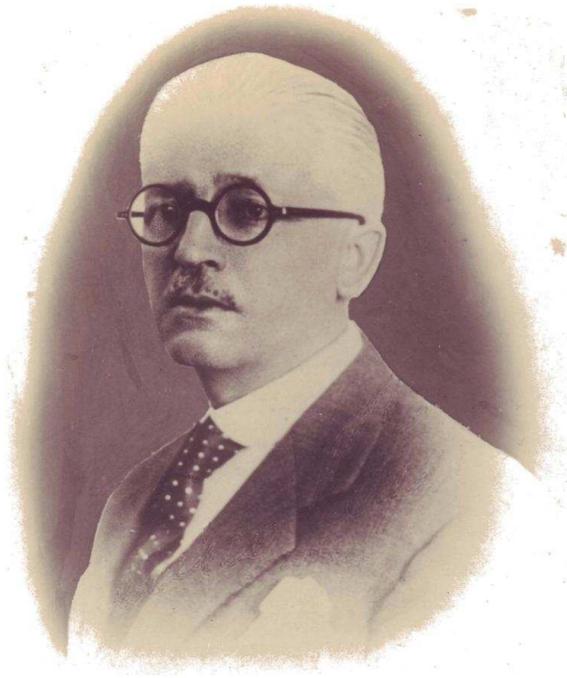} 
\caption{Domenico Pacini, picture taken from the obituary in Bari
University.}
\label{fig:pacini}
\end{figure}

The conclusion
that radioactivity was mostly coming from the Earth's crust was questioned by
the Italian physicist Domenico Pacini, who compared the rate of ionization on
mountains, over a lake, and over the sea \cite{Pac1909,Pac1910}; in 1911, he
made important experiments by immersing an electroscope deep in the sea
\cite{Pac1912}. 

Pacini\cite{Riz1934} (Fig. \ref{fig:pacini}) was born in 1878,
in Marino, near
Roma. He graduated in Physics in 1902 at the Faculty of Sciences of Roma 
University. There, for the next three years, he worked as an assistant to
Professor Blaserna studying electric conductivity in gaseous media. In 1906
Pacini was appointed assistant at Italy's Central Bureau of Meteorology and
Geodynamics, heading the department that was in charge of studying thunderstorms
and electric phenomena in the atmosphere. Pacini held that position until 1927,
when he was upgraded to Principal Geophysicist. Finally in 1928 he was appointed
full professor of Experimental Physics at the University of Bari. Pacini died of
pneumonia in Roma  in 1934, shortly after his marriage.

Pacini's important results on the penetrating radiation started with studies on
electric conductivity in gaseous media performed at the University of Roma 
during the early years of the 20th century.  During 1907-1912, he performed
several detailed measurements on the conductivity of air, using an Ebert
electroscope to enhance the sensitivity (he could reach a
sensitivity of one third of volt).

\begin{figure}[hb]

\includegraphics[height=15cm,width=15cm]{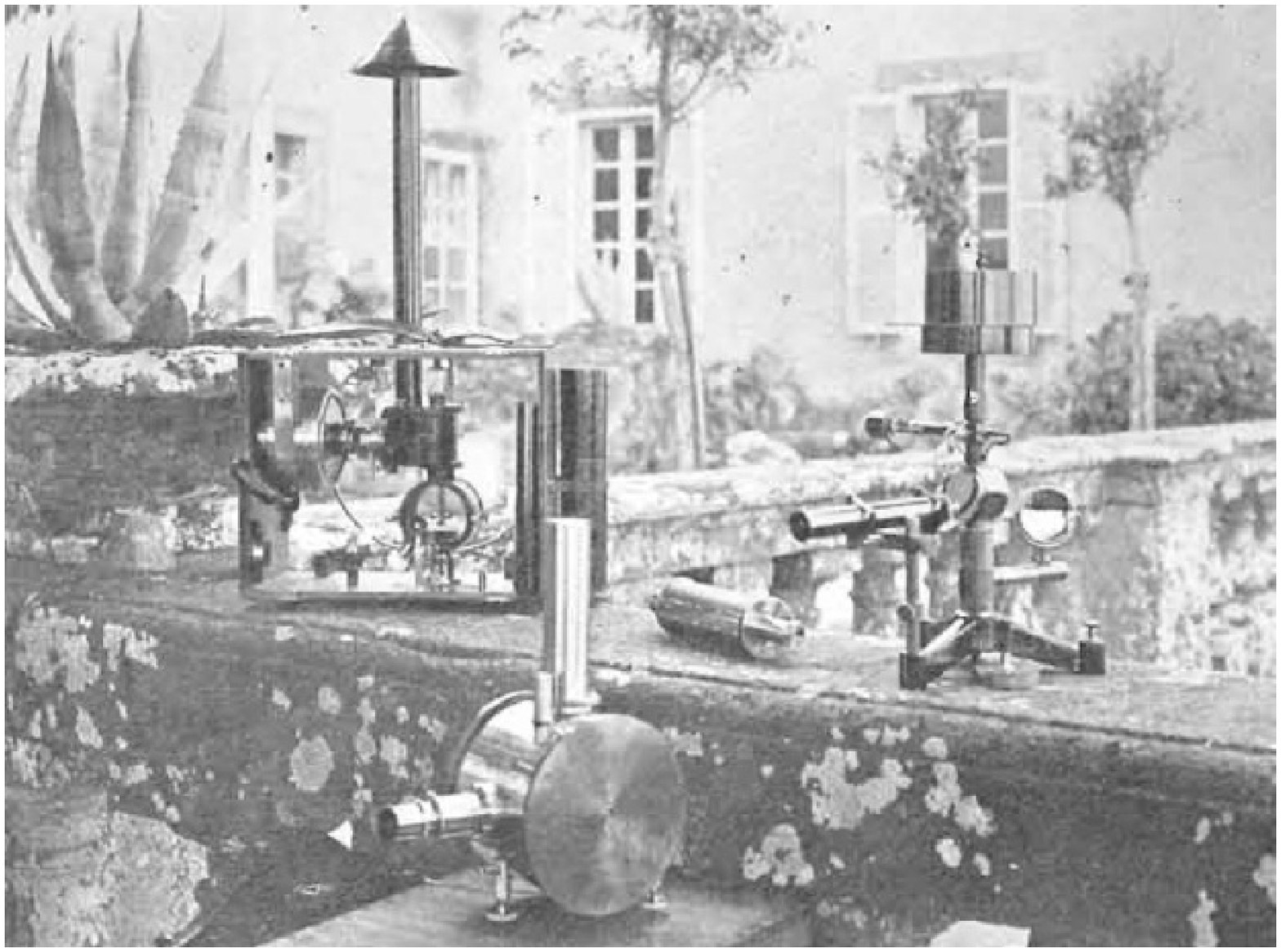} 
\caption{Pacini's electroscopes, picture included in  \cite{Pac1912}.}
\label{fig:electroscopes}
\end{figure}

 In a first period Pacini made several measurements to establish the variations
of the electroscope's discharge rate as a function of the environment. First he
placed the electroscope on the ground and on the sea a few kilometers off the
coast; the results were comparable. An example of the electroscopes used by
Pacini is in Fig.\ref{fig:electroscopes} taken from the paper in
\cite{pacini08}.

A summary of these results indicates,
according to Pacini's conclusions, that ``in the hypothesis that the origin of
penetrating radiations is in the soil, since one must admit that they are
emitted at an almost constant rate (at least when the soil is not covered by
remaining precipitations), it is not possible to explain the results obtained''
\cite{Pac1909}.

 It is clear that Pacini's conclusion, confirmed by Gockel
\cite{Goc1909}, was the first that clearly affirmed the results of many
experiments on radiation could not be explained by the radioactivity in the
Earth's crust. 

\begin{figure}
\includegraphics[height=9cm,width=8cm]{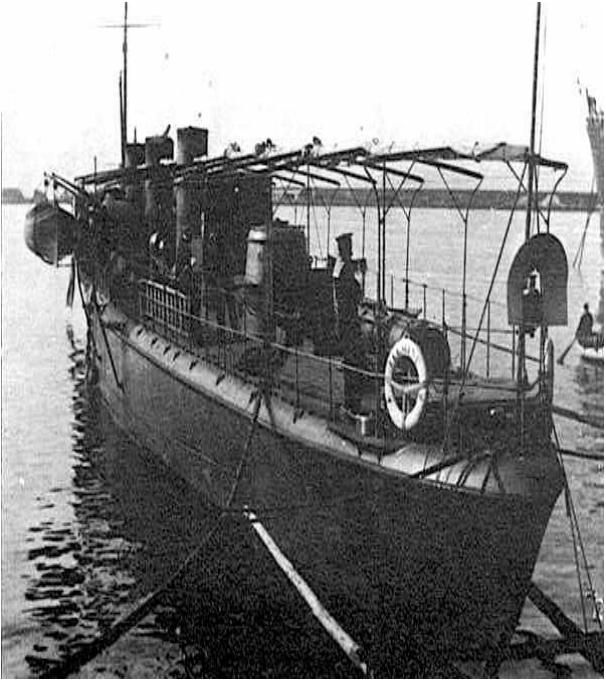}
\caption{The cacciatorpediniere ``Fulmine'', used by Pacini for his measurements
on the sea.}
\label{fig:fulmine}       
\end{figure}

Pacini continued the investigations of radiation and developed in
1911 an experimental technique for underwater measurements\cite{Pac1912,Pac11}. 
From this point of view
it's evident that Pacini was among the first scientists to think and use
``underground-undersea'' measurements to study what later we called Cosmic Rays.
He found a significant
decrease in the discharge rate when the electroscope was placed underwater.
Pacini made his measurements over the sea in the Gulf of Genova\cite{Pac11}, on
an Italian
Navy ship, the {\em cacciatorpediniere} (destroyer)  ``Fulmine" (Fig.
\ref{fig:fulmine}) from the Accademia Navale di Livorno. 

He reported those measurements,  results, and their interpretation,
in a note titled {\em La radiazione penetrante alla superficie ed in seno alle
acque (Penetrating radiation at the surface of and in water)} \cite{Pac1912}. In
that paper Pacini wrote: ``Observations carried out on the sea during the year
1910 \cite{Pac1910} led me to conclude that a significant proportion of the
pervasive radiation that is found in air had an origin that was independent of
direct action of the active substances in the upper layers of the Earth's
surface. ... [To prove this conclusion] the apparatus ... was enclosed in a
copper box so that it could immerse in depth. ... From June 24 to June 31 [sic!]
[1911], observations were performed with the instrument at the surface, and with
the instrument immersed in water, at a depth of 3 meters.''  

With the apparatus
at the surface 300 m from land, Pacini measured seven times during three hours
the discharge of the electroscope, obtaining a loss of 12.6 V/hour,
corresponding to 11.0 ions per second (with a RMS of 0.5 V/hour); with the
apparatus at a 3 m depth in the 7 m deep sea, he measured an average loss of
10.3 V per hour, corresponding to 8.9 ions per second (with a RMS of 0.2 V/h).
Consistent results were obtained during measurements at the Lake of Bracciano.

 The measurement
underwater was 20\% lower than at the surface, consistent with absorption by
water of a radiation coming from above. ``With an absorption coefficient
of
0.034 for water, it is easy to deduce from the known equation $I/I_0$ =
exp(-d/$\lambda$), where d is the thickness of the matter crossed, that, in the
conditions of my experiments, the activities of the sea-bed and of the surface
were both negligible. 
 The explanation appears to be that, due to the absorbing
power of water and the minimum amount of radioactive substances in the sea,
absorption of radiation coming from the outside happens indeed, when the
apparatus is immersed.'' Pacini concluded \cite{Pac1912}: ``[It] appears from
the results of the work described in this Note that a sizable cause of
ionization exists in the atmosphere, originating from penetrating radiation,
independent of the direct action of radioactive substances in the soil." 

By the way, in 1910 Pacini\cite{pacini08} looked for a possible increase in
radioactivity during a passage of the Halley's comet (and he found no effect of
the comet itself). 
Similar sea measurements were performed by Simpson and Wright in
1911\cite{simpson11}  showed a relevant ionization over the sea, a result
not accounted by the soil radioactivity, since sea water should contain only a
minor fraction of radioactive elements.

\subsection{Hess and the balloon measurements}

The need for balloon experiments became evident to clarify Wulf's observations
on the effect of altitude (at that time and since 1885, balloon experiments were
anyway widely used for studies of the atmospheric electricity). The first
balloon flight with the purpose of studying the properties of penetrating
radiation was arranged in Switzerland in December 1909 with a balloon called
Gotthard from the Swiss aeroclub. Alfred Gockel, professor at the University of
Freiburg, ascending in a balloon up to 4500 m above sea level (a.s.l.) during
three successive flights, found \cite{Goc1910,Goc1911} that the ionization did
not decrease with height as expected on the hypothesis of a terrestrial origin.
Gockel confirmes the conclusion of Pacini and concludes ``that a non-negligible
part of the penetrating radiation is independent of the direct action of the
radioactive substances in the uppermost layers of the Earth'' \cite{Goc1911}. 
In
spite of Pacini's conclusions, and of Wulf's and Gockel's puzzling results on
the dependence of radioactivity on altitude, physicists were however skeptical
about the hypothesis of a non-terrestrial origin. 
\begin{figure}[htb]
\vspace{12pt}
\includegraphics[height=10cm,width=7cm]{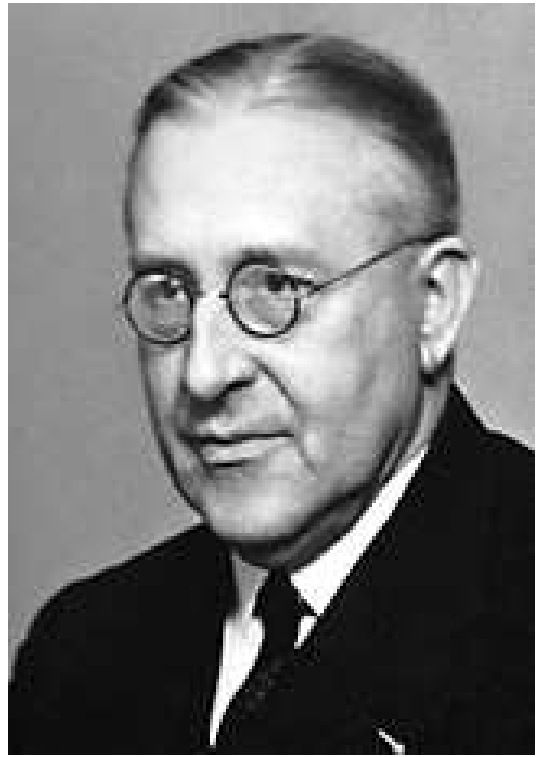}
\caption{Victor Franz Hess.}
\label{fig:hess}
\end{figure}

The situation was cleared up
thanks to a long series of balloon flights by the Austrian physicist Victor
Hess (Fig.\ref{fig:hess}), who established the extra-terrestrial origin of at
least part of the
radiation causing the observed ionization. 

Hess was born in 1883 in Steiermark,
Austria, and he took his doctor's degree in 1910 in Graz. After graduation he
was assistant under professor Meyer at the Institute of Radium Research of the
Viennese Academy of Sciences, where he performed most of his work on cosmic
rays, and in 1919 he became Professor of Experimental Physics at the Graz
University. Hess was on leave of absence from 1921 to 1923 and worked in the
United States, where he took a post as Director of the Research Laboratory of
the United States Radium Corporation, at Orange (New Jersey). In 1923 he
returned to Graz and in 1931 he moved to Innsbruck as professor. In 1936 Hess
was awarded the Nobel Prize in physics for the discovery of Cosmic Rays. After
moving to USA in 1938 as professor at Fordham University, Hess became an
American citizen in 1944, and lived in New York until his death in 1964. Hess
started his experiments by studying Wulf's results, and knowing the detailed
predictions by Eve\cite{Eve1911} on the coefficients of absorption for
radioactivity in the atmosphere. Eve wrote that, if one assumed a uniform
distribution of RaC on the surface and in the uppermost layer of the Earth, ``an
elevation to 100 m should reduce the [radiation] effect to 36 percent of the
ground value''. Hess added: ``This is such a serious discrepancy [with Wulf's
results] that its resolution appears to be of the highest importance for the
radioactive theory of atmospheric electricity'' \cite{Hes1912}. Since in the
interpretation of Wulf's and Gockel's results the absorption length of the
radiation (at that time identified mostly with gamma radiation) in air entered
crucially, Hess decided first to improve the experimental accuracy of the Eve's
result by ``direct measurements of the absorption of gamma rays in air''
\cite{Hes1911}. He used  probes of about 1 g RaCl$_2$ at distances up to 90 m,
and obtained an absorption coefficient consistent with Eve. Hence the
contradiction of Wulf's results remained; Hess concluded that ``a clarification
can only be expected from further measurements of the penetrating radiation in
balloon ascents'' \cite{Hes1911}.
\begin{figure}[h]
\includegraphics[width=0.95\columnwidth,height=9cm]{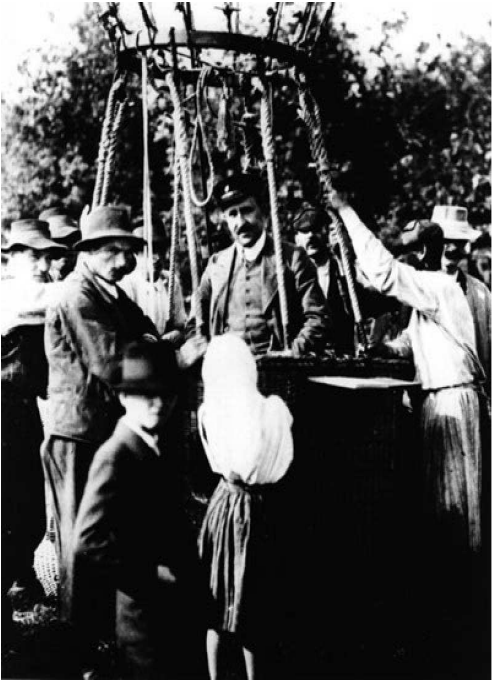}
\caption{Historical photograph of Hess preparing for a balloon flight.}
\label{fig:balloon}       
\end{figure}

 Hess continued his studies with balloon
observations (Fig. \ref{fig:balloon}). The first ascension took place on August
28, 1911. ``[T]he balloon" Radetzky" of the Austrian aeroclub  with
Oberlutnant
S. Heller as pilot and me as sole passenger was lifted'' \cite{Hes1911}. The
ascension lasted four hours and went up to a height of 1070 m above ground. A
second ride in another balloon (``Austria'') during the night of 12 October 1911
went up to 360 m above ground. During both balloon flights, the intensity of the
penetrating radiation was measured to be constant with altitude within errors.
From April 1912 to August 1912 Hess had the opportunity to fly seven times with
three instruments (two with thick walls and one with thin walls, to disentangle
the effect of beta radiation). In the final flight, on August 7, 1912, he
reached 5200 m. To his surprise, the results clearly showed that the ionization,
after passing through a minimum, increased considerably with height. 

 ``(i)
Immediately above ground the total radiation
decreases a little. (ii) At altitudes of 1000 to 2000 m there occurs again a
noticeable growth of penetrating radiation. (iii) The increase reaches, at
altitudes of 3000 to 4000 m, already 50\% of the total radiation observed on the
ground. (iv) At 4000 to 5200 m the radiation is stronger [more than 100\%] than
on the ground'' \cite{Hes1912}.

Hess concluded that the increase of the ionization with height must be due to a
radiation coming from above, and he thought that this radiation was of
extra-terrestrial origin. He also excluded the sun as the direct source of this
hypothetical penetrating radiation because of no day-night variation. Hess
finally published a summary of his results in Physikalische Zeitschrift in 1913
\cite{Hes1913}, a paper which reached the wide public. 

The results by Hess were
later confirmed by Kolh\"orster \cite{Kol1914} in a number of flights up to 9200
m.  An increase of the ionization up to ten times that at sea level was found.
The absorption coefficient of the radiation was estimated to be 10$^{-5}$ per cm
of air at NTP. This value caused great surprise as it was eight times smaller
than the absorption coefficient of air for gamma rays as known at the time.
Kolh\"orster continued his investigations using newly constructed apparatuses
and made measurements at mountain altitudes with results published in
1923\cite{Kol1923} in
agreement with earlier balloon flights. There were, however, also negative
attitudes in Europe against an extra-terrestrial radiation. Hoffmann, using
newly developed electrometers, concluded \cite{hoff} that the cause of the
ionisation was radioactive elements in the atmosphere. Similar conclusions were
reached by Behounek \cite{beh} and for several years this convinction was no
more discussed.

\section{Toward the final understanding of the  origin of the
radiation}
After the war, the focus of the research moved to the US. Millikan and Bowen
\cite{Mil1923} developed a low mass (about 200 g) electrometer and ion chamber
for unmanned balloon flights using data transmission technology developed during
World War I. In balloon flights to 15,000 m in Texas they were surprised to find
a radiation intensity not more than one-fourth the intensity reported by Hess
and Kolh\"orster. They attributed this difference to a turnover in the intensity
at higher altitude, being unaware that a geomagnetic effect existed between the
measurement in Europe and Texas. Thus, Millikan believed that there was no
extraterrestrial radiation. As reported to the American Physical Society in 1925
Millikan's statement was ``The whole of the penetrating radiation is of local
origin''.

In 1926, however, Millikan and Cameron \cite{Cam1926} carried out absorption
measurements of the radiation at various depths in lakes at high altitudes.
Based upon the absorption coefficients and altitude dependence of the radiation,
they concluded that the radiation was high energy gamma rays and that ``these
rays shoot through space equally in all directions'' calling them ``cosmic
rays''. 

Millikan was handling with energy and skill the communication with the media,
and in the US the discovery of Cosmic Rays became, according to the public
opinion, a success of American science. Millikan argued that the radiations are
``generated by nuclear changes having energy values not far from [those that
they recorded] in nebulous matter in space.'' Millikan then proclaimed that this
cosmic radiation was the ``birth cries of atoms'' in our galaxy. His lectures
drew considerable attention from, among others, Eddington and Jeans, who
struggled unsuccessfully to describe processes that could account for Millikan's
claim.
It was generally believed that the cosmic radiation was gamma radiation because
of its penetrating power (one should remember that the penetrating power of
relativistic charged particles was not known at the time). Millikan had put
forward the hypothesis that the gamma rays were produced when protons and
electrons form helium nuclei in the interstellar space. 

A key experiment, which would decide the nature of Cosmic Rays (and in
particular if they were charged or neutral), was the measurement of the
dependence of cosmic ray intensity on geomagnetic latitude. Important
measurements were made in 1927 and 1928 by Clay \cite{clay} who, during two
voyages between Java and Genova, found that the ionisation increased with
latitude. No such variation was expected if the radiation was a gamma radiation,
but Clay did not draw a firm conclusion as to the nature of the cosmic
radiation. Clay's work was disputed by Millikan.

With the introduction of the Geiger-Muller counter in 1928, a new era began and
soon confirmation came that the cosmic radiation is indeed corpuscular.
Kolh\"orster introduced the coincidence technique. Bothe and Kolh\"orster
\cite{BK} concluded that the cosmic radiation is mainly or fully corpuscular,
but still Millikan did not accept this view. 

In 1932 Compton carried out a world-wide survey to settle the dispute.  He then
reported \cite{Com1933} that there was a latitude effect, that Cosmic Rays were
charged particles and that Millikan was wrong.  Millikan attacked strongly
Compton, but after repeating his experiment in 1933 he admitted that there was a
latitude effect and that the Cosmic Rays must be (mostly) charged particles.
However, it would take until 1941 before it was established in an experiment by
Schein \cite{sche} that Cosmic Rays was mostly protons.

The  1936 Nobel Prize in Physics was shared by professor V.F. Hess for the
discovery of Cosmic Rays  and dr. C.D. Anderson for the discovery of the
positron.

Professor
H. Pleijel, Chairman of the Nobel Committee for Physics of the Royal
Swedish Academy
of Sciences, said in his speech at the Nobel award ceremony in 1936:
"[A] search was made throughout nature for radioactive substances [by
several scientists]:
in the crust of the Earth, in the seas, and in the atmosphere; and
... the electroscope was applied. Radioactive rays were found everywhere,
whether investigations were made into the waters of deep lakes, or
into high mountains.
... Although no definite results were gained from these
investigations, they did show that
the omnipresent radiation could not be attributed to radiation of
radioactive substances
in the Earth's crust...The mystery of the origin of this radiation remained
[however] unsolved until Prof. Hess made it his problem. ... With
superb experimental
skill Hess perfected the instrumental equipment used and eliminated
its sources of error.
With these preparations completed, Hess made a number of balloon ascents... From
these investigations Hess drew the conclusion that there exists an
extremely penetrating
radiation coming from space which enters the Earth's atmosphere.

\section{Conclusions}
The history of Cosmic Rays discovery may be divided into two era:
in the early stage a lot of people contributed to the understanding of the
origin and the characterization of the radiation as penetrating radiation.
Pacini, Wulf,  Gockel and Hess among the others, contributed both to
the developement of the instruments to detect the effects and characterize the
unknown penetrating radiation.

Moreover this people identified the basic three ways to study Cosmic Rays, i.e.
underground-undersea measurements, ground and balloon-borne or space
measurements, techniques that really were applied only about 40 years later when
the detectors evolved towards the particle physics.
In every case the conclusion by the Royal Swedish Academy of Sciences can be
repeated: the
discovery of Cosmic Rays has opened new areas for experimental and theoretical
physics of greatest significance for our understanding of the structure and
origin of matter and that Hess measurements established clearly that the
radiation  had an extraterrestrial origin.

We hope that this note will establish correctly the effort of many people
involved in the discovery of Cosmic Rays about 100 years later the most
relevant set of measurements that removed the ambiguity about the origin: in
particular D. Pacini with his underground measurements and V.Hess with  his
balloon measurements, have clearly excluded the soil radioactivty as the main
reason for the penetrating radiation, and posed the basis for the moder cosmic
ray physics.

\section*{Acknowledgements}
We are grateful to the University of Bari, and in particular to Professor
Augusto Garuccio, 
for supporting the research of documents regarding Domenico Pacini; to the
Dipartimento 
Interateneo di Fisica of Bari for jointly organizing the Domenico Pacini
memorial day that 
was held in Bari on April 17, 2007; to Professors Guerra and Robotti for
uncovering relevant 
material in the Amaldi Archive at Rome's ''La Sapienza'' University and in the
Bracciano 
museum,  to A.~De~Angelis and P.Carlson \cite{CarlsonDeangelis} and
S.~Stramaglia for their effort and support to translate and find new documents 
the origin Pacini's
papers and to their contribution to this work.

\begin{thebibliography}{9}
\newcommand{\BY}[1]{{#1},}
\newcommand{\IN}[4]{{#1} \textbf{#2} (#3) #4}

\bibitem{Wolfendale1984}
  \BY{A.~W.~Wolfendale},
  Rept.\ Prog.\ Phys.\  {\bf 47} (1984) 655.
\bibitem{rossi} \BY{B. Rossi} {\em Cosmic Rays} (McGraw-Hill, New York 1964)
\bibitem{hillas} \BY{A.M. Hillas} {\em Cosmic Rays} (Pergamon Press, Oxford
1972)
\bibitem{wilson} \BY{J.G. Wilson} {\em Cosmic Rays} (Chapman and Hall, London
1976)
\bibitem{janossy} \BY{L. J\'anossy}  {\em Cosmic Rays} (Clarendon
Press, Oxford
1950)
\bibitem{xu}\BY{Q. Xu and L.M. Brown} \IN{Am. J. Phys.}{55}{1987}{23}



\bibitem{rob}F. Guerra e N. Robotti, ``La scoperta dei raggi cosmici: Domenico
Pacini'', lezione alla Scuola di Dottorato di Otranto  (2007), unpublished

\bibitem{stra1} \BY{A. De Angelis, N. Giglietto, L. Guerriero, E. Menichetti, P.
Spinelli, S. Stramaglia} \IN{Nuovo Sagg.}{24}{2008}{70}
\bibitem{stra2} \BY{A. De Angelis, N. Giglietto and S. Stramaglia} {\em 
Domenico Pacini, the forgotten pioneer of the discovery of Cosmic Rays,} 
arXiv:1002.2888 (February 2010)

\bibitem{CarlsonDeangelis} P. Carlson and A. De Angelis, 
  P.~Carlson and A.~De Angelis,
  arXiv:1012.5068 [physics.hist-ph].
accepted for pubblication in European Physical Journal H.

\bibitem{Coulomb} \BY{C. de Coulomb} {\em Mdm. de l'Acad. des Sciences} (Paris,
1875) p. 612


\bibitem{Bequerel} \BY{H. Becquerel}  {\em Sur les radiations \'emises par
phosphorescence (On the radiation emitted by phosphorescence),} 
\IN{Comptes Rendus de l'Acad. des Sciences}{122}{1896} {420}


\bibitem{Wilson1901} \BY{C.T.R. Wilson} {\em On the Ionisation of Atmospheric
Air,}\IN{Proc. Roy. Soc. of London}{68}{1901}{151}

\bibitem{Geitel} \BY{J. Elster and H.F. Geitel} \IN{Ann. d.
Phys.}{2}{1900}{425}

\bibitem{Rutherford} \BY{E. Rutherford and H.L. Cook} \IN{Phys.
Rev.}{16}{1903}{183}
\bibitem{burton} \BY{F.C. McLennan and F. Burton} \IN{Phys.
Rev.}{16}{1903}{184}

\bibitem{kurz} \BY{K. Kurz} \IN{Phys. Zeit.}{10}{1909} {834}

\bibitem{Cline1910} Cline, G.~A.\ 1910, Physical 
Review Series I, 30, 35
\bibitem{Borgman1905} Borgman J.J., Science Abstracts, 1580,(1905).
\bibitem{mach1906} Mach and Rimmer, Phys.Zeit.,7, (1906),617
\bibitem{pacini1909} Pacini, D., Rend. Acc.Lincei, 18, (1909),123
\bibitem{Riz1934} \BY{G.B. Rizzo} {\em Domenico Pacini (1878-1934),} \IN{Nuovo
Cim.}{11}{1934}{509};

\bibitem{Wul1910} \BY{Th. Wulf} \IN{Phys. Zeit.} {1}{1909}{152}
\bibitem{Goc1909}   \BY{A. Gockel} \IN{Phys. Zeit.} {10}{1909} {845}
\bibitem{eve} \BY{A.S. Eve} \IN{Philos. Mag.} {13}{1907}{248}
\bibitem{Wul1907} \BY{Th. Wulf} \IN{Phys. Zeit.}{8}{1907}{246,527 and 780}
\bibitem{Pac1909} \BY{D. Pacini} \IN{Rend. Acc. Lincei} {18}{1909} {123}
\bibitem{Pac1910} \BY{D. Pacini}  \IN{Ann. Uff. Centr. Meteor.} {XXXII, parte
I}{1910}{};  \IN{Le Radium} {VIII}{1910} {307}
\bibitem{Pac1912} \BY{D. Pacini} 
\IN{Nuovo Cim.} {VI/3} {1912} {93}, translated and commented by A. De Angelis,
{\em Penetrating radiation at the surface of and in water,} (arXiv:1002.1810v1
[physics.hist-ph])

\bibitem{Pac11}D. Pacini, ``La radiazione penetrante sul mare'', Ann. Uff.
Centr. Meteor. XXXII, parte I (1910); 
``La radiation penetrante sur la mer'' Le Radium VIII (1911) 307

T. VIII, pag. 307, 1911
\bibitem{pacini08}D. Pacini, ``Misure di ionizzazione dell'aria su terraferma
ed 
``Sulla radioattivit\`a indotta dell'atmosfera nel golfo ligure'', Nuovo Cim.
V/15 (1908) 24; 
``Questioni di elettricit\`a atmosferica'', Nuovo Cim. V/19 (1910) 449; 
``Sui prodotti del radio e del torio nell'atmosfera'', Nuovo Cim. V/19  (1910)
345; 
``Osservazioni di elettricit\`a atmosferica eseguite in occasione del
passaggio %
della cometa di Halley'', Ann. Uff. Centr. Meteor. XXXII, parte I
(1910)
\bibitem{simpson11} \BY{Simpson, G. C. and Wright} \IN{Proc. Roy.
Soc.}{85}{1911}{175}
\bibitem{Goc1910} \BY{A. Gockel} \IN{Phys. Zeit.} {11}{1910} {280}

\bibitem{Goc1911}   \BY{A. Gockel} \IN{Phys. Zeit.} {12}{1911} {595}

\bibitem{Eve1911} \BY{A.S. Eve} \IN{Philos. Mag.} {21} {1911}{27}

\bibitem{Hes1912} \BY{V. Hess} \IN{Phys. Zeit.} {13}{1912} {1084}
\bibitem{Hes1911} \BY{V. Hess} \IN{Phys. Zeit.} {12}{1911} {998} 

\bibitem{Hes1913} \BY{V. Hess} \IN{Phys. Zeit.} {14}{1913} {612}
\bibitem{Kol1914} \BY{W. Kolh\"orster} \IN{Phys. Zeit.} {14} {1913}{1153};
\IN{Ber. deutsch. Phys. Qes.}{16}{1914}{719}

\bibitem{Kol1923}W. Kolh\"rster, \IN{Berlin. Ber.}{34} {1923}{}
\IN{Phys. Zeit.} {25}{1924} {177};  \IN{Phys. Zeit.} {26}{1925} {40 and 669} 

\bibitem{hoff} \BY{G. Hoffmann} \IN{Zeit. f. Phys.} {7}{1921} {254}; \IN{Phys.
Zeit.} {24}{1923} {475};  
\bibitem{beh} \BY{F. Behounek} \IN{Phys. Zeit.} {27} {1926}{8} 
\bibitem{Mil1923} \BY{R.A. Millikan and I.S. Bowen} \IN{Phys. Rev.} {22}{1923}
{198}
\bibitem{Cam1926} \BY{R.A. Millikan and G.H. Cameron} \IN{Phys. Rev.} {28}{1926}
{851}
\bibitem{clay} \BY{J. Clay} \IN{Acad. Amsterdam Proc.}{30} {1927}{1115}; 
\IN{Acad. Amsterdam Proc.}{31} {1928}{1091}
\bibitem{BK} \BY{W. Bothe and W. Kolh\"orster} \IN{Zeit. f. Phys.} {56}{1929}
{751}
\bibitem{Com1933} \BY{A. Compton} \IN{Phys. Rev.} {41}{1932} {681};  \IN{Phys.
Rev.} {43}{1933} {387} 

\bibitem{sche} \BY{M. Schein et al.} \IN{Phys. Rev.} {59} {1941} {615}









\end{thebibliography}
\end{document}